\documentclass[final,oneside]{igs}
\usepackage{igsnatbib}
\usepackage{float}

\ifx\pdftexversion\undefined
\usepackage[dvips]{graphicx}
\else
\usepackage[pdftex]{graphicx}
\usepackage{epstopdf}
\usepackage{hyperref}
\usepackage{amsmath}
\fi

\begin{document}

\title[Sea Ice Brine Fraction]{The Sensitivity of Sea Ice Brine Fraction to the Freezing Temperature and Orientation}

\author[Stewart et al.]{Kial~D.~STEWART,$^{1,2}$
  William~PALM,$^{1,2}$
  Callum~J.~SHAKESPEARE,$^{1,2}$ 
  Noa~KRAITZMAN$^3$}

\affiliation{%
$^1$Climate \& Fluid Physics Laboratory, Australian National University\\
$^2$Australian Research Council Centre of Excellence for Climate Extremes\\
$^3$Macquarie University\\
  Correspondence: Kial Stewart
  $<$kial.stewart@anu.edu.au$>$}

\abstract{Pound for pound, sea ice is the most important component of Earth's climate system.
The changing conditions in which sea ice forms and exists are likely to affect the properties of sea ice itself, and potential climate feedbacks need to be identified and understood to improve future projections.
Here we perform a set of idealised experiments which model sea ice growth under a range of ambient salinities, freezing temperatures, and freezing orientations.
The results confirm existing theories; sea ice growth rate is largest for cooler freezing temperatures, fresher ambient salinities, and bottom oriented freezing configuration.
Our primary metric of interest is the brine mass fraction (the mass ratio of brine inclusions to the total sea ice), which we quantify and determine the sensitivity of with respect to changes in freezing conditions.
We find that the brine mass fraction of our model sea ice is most sensitive to freezing temperature, and increases 2.5\% per 1$^\circ$C increase of freezing temperature.
This finding suggests that future sea ice in warmer climate states will retain more brine, with subsequent flow on effects for circulations driven by brine rejection in the high-latitude oceans.}

\maketitle
\section{Introduction}

Sea ice is a vital component of Earth's climate system, covering more than 10\% of the global oceans at least part of each year \citep{weeks2010, eayrs_etal2019}.
Its high albedo reflects between 80-90\% of the incoming solar radiation, thereby strongly limiting the radiative heating of polar surface waters.
Brine rejection from sea ice increases the salinity of high-latitude surface ocean waters that ultimately sink and play a key role in the global thermohaline overturning circulation.
Sea ice is also highly sensitive to changes in climate, and is currently exhibiting repeated record low areal coverage \citep[e.g.,][]{purich_doddridge2023}, which serves as an important measure of how the Earth system is responding to climate change.

Despite the importance of its role in Earth's climate, and its unique set of dynamics and thermodynamics, the representation of sea ice in coupled climate models tends to mirror that of the oceanic circulation models.
That is, the sea ice field is discretised onto the ocean's horizontal grid, the sea ice state and characteristics are represented with bulk values, and the dynamics/thermodynamics are approximated with a set of continuous partial differential equations that are solved in time.
Theories and limited existing observations are used to develop parameterisations of small scale dynamics that relate sub-grid scale processes to the larger scales that are resolved by the horizontal grid.
Many dynamics specific to sea ice, in particular its rheology, need to be omitted because they are not well represented by the ocean's horizontal grid.
Nevertheless, sea ice model development continues to improve and include additional processes, whose representation in turn often require new parameterisations.
For example, the latest version of the CICE sea ice model (CICE6 v6.1.2) includes a capability for a floe size distribution, which then permits lateral sea ice growth/melt dynamics, which requires specific parameterisations that are distinct from vertical sea ice growth/melt.
This new feature raises the question of whether the direction of sea ice growth changes the bulk characteristics of the sea ice itself \citep[e.g.,][]{scotti_etal2019}.

The difficulties associated with the realistic representation of sea ice in climate models are a major scientific challenge for predicting future climate states.
Sea ice is a particularly sensitive entity in the climate system and prone to feedback cycles, as demonstrated by the ice-albedo phenomenon \citep[e.g.,][]{budyko1969}.
The polar amplification of global warming \citep[e.g.,][]{manabe_stouffer1979} will tend to the warm surface atmosphere at higher latitudes faster than lower latitudes; this will result in sea ice growing in conditions that are relatively warmer than before.
Understanding how sea ice properties respond to warmer freezing temperatures is an important step for improving sea ice and climate models.

Here, we investigate the impacts of freezing temperature and orientation on the brine mass fraction of sea ice by conducting a suite of laboratory experiments to model sea ice growth across a range of freezing temperatures and orientations.
The bulk brine mass fraction of the model sea ice is estimated using two independent approaches, and found to be sensitive to both the freezing temperature and orientation across a wide range of ambient salinities.
In \S2 we provide background context for our study and subsequent analysis.
In \S3 we describe the laboratory apparatus, methodology and analysis, and present and discuss the results of the experiments in \S4.
We provide a brief commentary on the geophysical implications of our results in \S5.

\section{Background}\label{background}

Seawater is a multi-component liquid wherein its constituents have different freezing points.
In such a system, the freezing process typically results in the formation of a porous mixture of solid and liquid inclusions known as a mushy layer \citep{wells_etal2019}.
Solidification at temperatures above the melting temperature of the constituents results in the solvent freezing whilst the solutes remain liquid.
In the case of an aqueous solution containing salts, like seawater, as solidification occurs, finger-like solid crystals known as dendrites grow in advance of the solidifying face.
These dendrites cannot incorporate the dissolved salts into their matrix and expel it into the solution between the solid crystals \citep{anderson_guba2020}.
This enriched saline solution is referred to as brine.
As the dendrites continue to grow and the ice face advances, some of the liquid brine becomes trapped within the solid ice, forming brine inclusions.
The mixture of solid ice and brine inclusions forms the mushy layer \citep{feltham_etal2006}.
This mixture is generally classified by its solid mass fraction, which is the ratio of the solid ice mass to the total mass of the mush; the solid mass fraction is related to the brine mass fraction (the ratio of the liquid brine mass to the total mushy layer mass, which is often employed in sea ice models) as they sum to unity.

\begin{figure}[!ht]
\begin{center}
\includegraphics[width=0.8\textwidth]{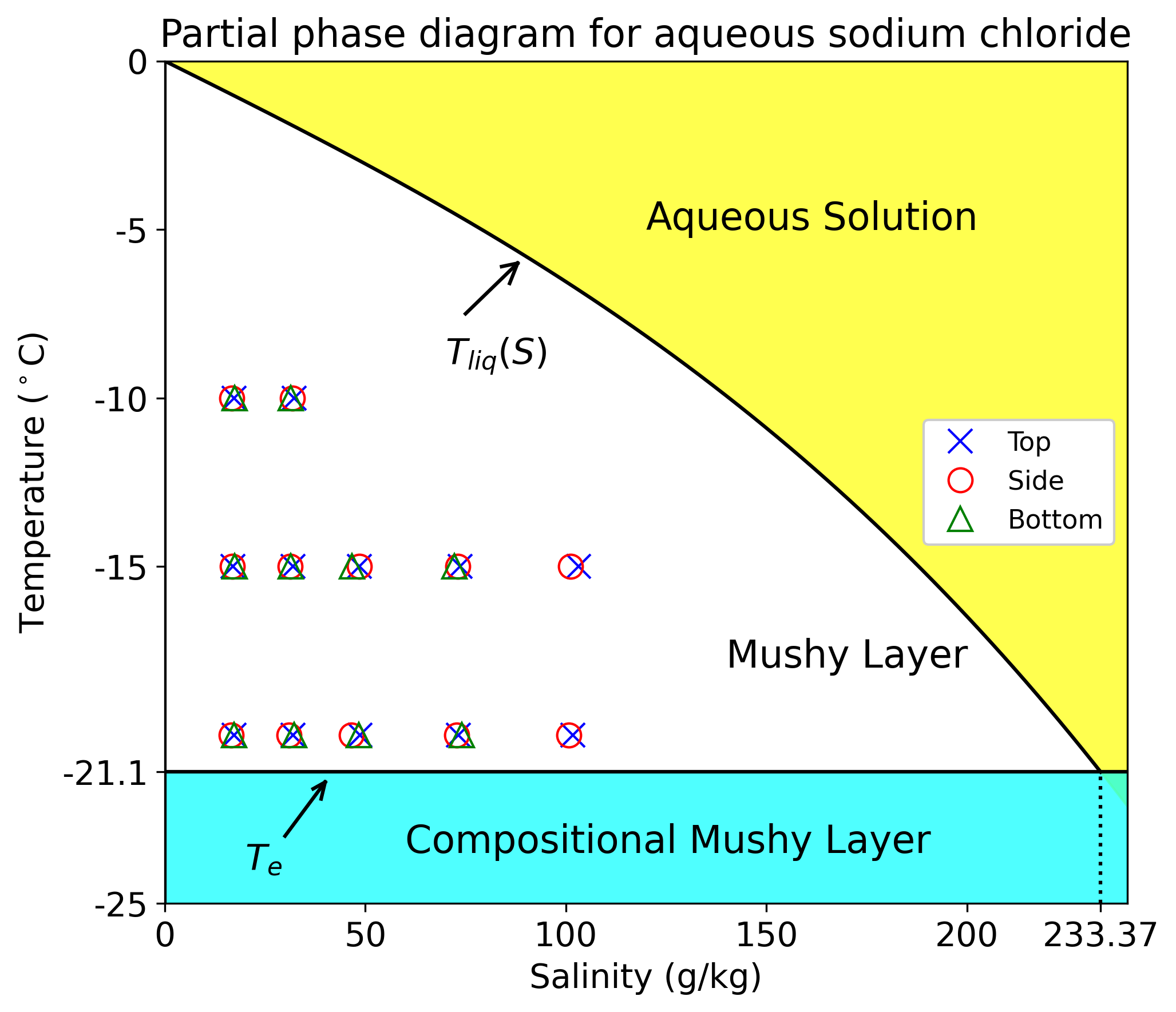}
\caption{A partial phase diagram for aqueous sodium chloride solutions. The blue region is cooler than the eutectic temperature $T_{e}=-21.1^{\circ}$C, at which sodium chloride crystal precipitate out of solution. The white region represents the mushy layer regime, and the yellow region represents the regime of aqueous solution wherein the salinity is too large for ice to form for the given temperature; these regimes are separated by the liquidus curve, which is given by $T_{liq}$ (Eqn. \ref{eqn:liquidus}) for temperatures warmer than eutectic temperature. The crosses, circles, and triangles represent the initial ambient salinities and freezing plate temperatures of the experiments with top, side, and bottom freezing plate orientations, respectively.}
\label{fig:parameter_space}
\end{center}
\end{figure}

Mushy layers are prevalent on Earth and form from both natural and industrial processes, such as the solidification of Earth's core, magma, and metallic alloys \citep{peppin_etal2007, huppert1990, anderson_guba2020}.
In the case of seawater, during the formation of sea ice, salts from the seawater are rejected from the ice lattice to form highly saline brine inclusions.
The high salinity of these brine inclusions results in them remaining in liquid phase within the ice; that is, in temperature--salinity space they exist to the right of the of the liquidus curve (e.g., Fig. \ref{fig:parameter_space}).

Mushy layers are subject to chemical and thermodynamic processes, and as such they react and evolve as the mushy layer grows \citep{wells_etal2019, anderson_guba2020}.
For example, brine inclusions must be in thermodynamic equilibrium within the ice.
If they are not, heat will flow to negate the temperature gradient such that thermodynamic equilibrium is established.
Therefore, due to the low temperatures within mushy layers, in order for the brine inclusions to remain liquid, they must exist on their liquidus curve, the solid-liquid phase transition point of a given substance.
This requires brine inclusions to vary their chemical and temperature characteristics as the mushy layer varies in temperature, which results in a reactive medium.

\cite{huppert1990} demonstrated that solidification of solutions with multiple dissolved salts below the eutectic temperature (the minimum temperature where all salts are able to be in solution) creates a compositional mushy layer; that is, an ice matrix which contains crystals of solid salt.
Here we avoid the added complexity of compositional mushy layers and focus on solidification at temperatures above the eutectic temperature for our aqueous solution.
We model the seawater and sea ice with a single-component liquid that is sodium chloride (NaCl) and fresh tapwater, which has a single eutectic temperature of $T_{e}=-21.1^{\circ}$C.
Note that actual seawater and sea ice, as a material composed of several different salts, does not have a eutectic point, and there is still liquid brine for temperatures below $-70^{\circ}$C, however precipitation begins at $T_{e}<-21.1^{\circ}$C \citep[e.g.,][]{weeks_ackley1986}, and modern sea ice models employ an equation of state with an imposed a eutectic temperature of $-36.2^{\circ}$C \citep[e.g.,][]{vancoppenolle_etal2018}.

The temperature difference between the relatively warmer ocean and relatively cooler atmosphere means that vertical temperature gradients exist within sea ice.
It follows that a similar temperature gradient exists within and surrounding individual brine inclusions \citep[e.g.,][]{kraitzman_etal2022}.
For a brine inclusion to maintain phase equilibrium, a salinity gradient is necessary within the liquid brine inclusions, which requires saltier, cooler liquid to exist at the top of the inclusion \citep{notz_worster2009}.
Obviously this arrangement is gravitationally unstable, and the brine inclusion itself is likely to be well-mixed.
That is, while the bulk-average of the brine inclusion is likely to be in thermal and chemical equilibrium, the upper and lower regions of the brine inclusion are not able to achieve equilibrium with their respective surrounds; the sea ice surrounding the upper regions of the inclusion will tend to freeze, while sea ice surrounding the lower regions will tend to melt \citep[e.g.,][]{weeks2010}.
This process allows the brine inclusions to tunnel through the sea ice which results in the brine inclusions travelling from the upper, cooler to the lower, warmer regions of the sea ice, and ultimately into the ocean below.
Without a source of salt in the upper sea ice region, the tunnelling of brine results in a monotonic decrease in the bulk sea ice salinity in time.
Additionally, the direction of the temperature gradient within the sea ice determines the tunnelling direction of brine inclusions; that is, horizontal gradients should result in horizontal tunnels.

The amalgamation of tunnelling brine inclusions can develop solid free chimneys referred to as brine channels \citep{worster1997}. 
Brine channels provide a link between the internals of the ice and the surrounding solution.
In the case of sea ice, the highly saline and near-freezing brine inclusions are much denser than the ocean waters immediately below.
Therefore, brine inclusions are negatively-buoyant and brine channels can facilitate convective fluid processes between sea ice and sea water \citep{worster_reesjones2015}.
Convective motion expels saline brine into the underlying ocean, in a process known as gravity drainage.
As brine channels provide a direct and active mechanism to drive flow between sea ice and the ocean, gravity drainage governs the rate of brine rejection.
Due to continuity, the volume of brine expelled from ice must be replaced, resulting in relatively less saline seawater replacing brine in the mushy layer \citep{wettlaufer_etal1997}.
As thermodynamic and chemical equilibrium must again be established, the less saline seawater in the ice freezes, re-initiating solidification and hence, increasing the solid mass fraction, and altering the mushy layer structure.
Increasing the solid mass fraction of ice decreases its permeability and thus a point is reached where convective motions can no longer be sustained, halting gravity drainage.
Therefore, gravity drainage is a mechanism for brine rejection that is most active during the early stages of sea ice growth.
Orientation of the freezing sea ice face will likely determine how effective gravity drainage is in facilitating brine rejection.

\section{Laboratory Experiments}\label{experiments}
\subsection{Apparatus}

\begin{figure}[!ht]
\begin{center}
\includegraphics[width=0.8\textwidth]{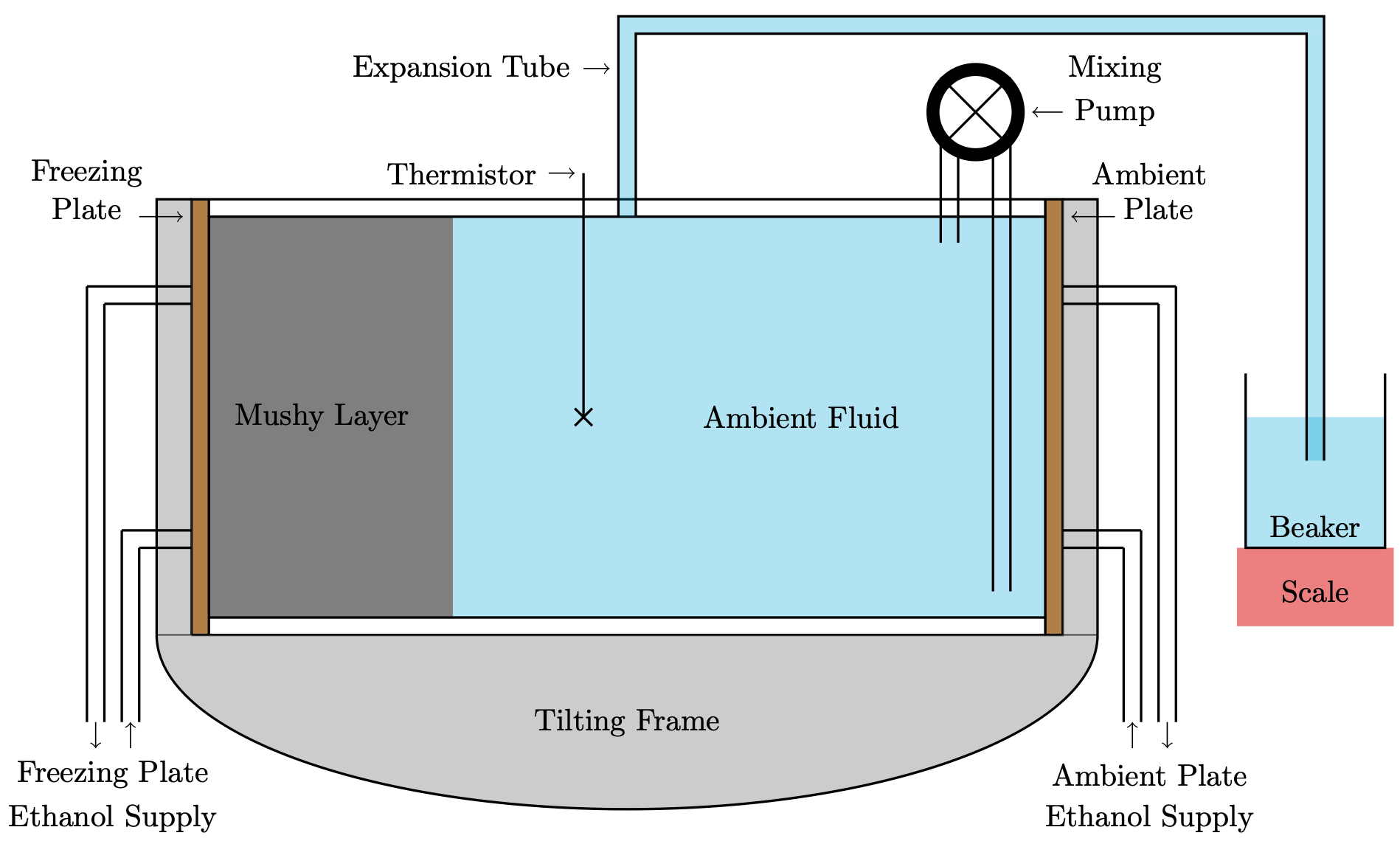}
\caption{A two-dimensional schematic of the experiment apparatus.}
\label{fig:exp_schematic}
\end{center}
\end{figure}

Our experiments investigating ice growth were conducted in the Climate \& Fluid Physics (CFP) laboratory at the Australian National University.
The experimental apparatus consists of a well-insulated, sealed, fluid-filled tank with internal dimensions $30.5{\times}21.5{\times}17.5$\,cm$^3$ (see Figure \ref{fig:exp_schematic}), of a similar design to that used by \cite{wettlaufer_etal1997}.
The two smallest area sidewalls of the tank are copper plates that are in direct thermal contact with controllable heat exchangers that allow the sidewall temperatures to be independently prescribed; the other four sidewalls are double-glazed perspex.
The tank is positioned between an illuminated white LED screen and a Basler AG camera that views the entire tank volume and configured such that it is able to clearly distinguish between the ambient fluid and the growing ice.
The apparatus is in a section of the CFP laboratory that is climate controlled and maintained at a constant temperature of $22.5\pm0.2^{\circ}C$.

Two independent Julabo FP50 HL refrigerating/heating constant temperature baths are used to prescribe the copper sidewall temperatures.
These constant temperature Julabo units pump ethanol through the sidewall heat exchangers, thereby providing a operational sidewall temperature range that easily spans from below the eutectic temperature of salty water ($T_{e}=-21.1^{\circ}$C) to above the freezing point of fresh water.
During operation, one of these copper sidewall plates is set to a temperature below the freezing point of the fluid, with the other plate used to maintain the temperature of the ambient fluid; these are referred to as the ``freezing" and ``ambient" plates, respectively.
So while the freezing plate initiates ice growth, the primary purpose of the ambient plate is to negate any unwanted heat flux into the tank from the laboratory, thereby maintaining the temperature of the ambient close to its freezing temperature.
The ambient fluid is kept well-mixed and homogeneous by an external pump that gently recirculates the ambient in the vicinity of the ambient plate and inhibits the formation of stratification.
The temperature of the ambient is logged using a thermistor positioned in the centre of the tank.

The effect of the freezing direction is one of the variables of interest.
For this, the tank is mounted in a semicircular frame that is able to be tilted such that the freezing plate can be oriented at the top, side, or bottom of the tank.

As ice grows against the freezing plate, there is an increase in total volume that is associated with the phase change from liquid to ice.
This volume increase is accommodated by a displacement of ambient fluid through an expansion tube and into a beaker that is on a scale, which records the measured mass in time.


\subsection{Methodology}

The tank was tilted in the desired orientation and the tank, expansion tube, and scale beaker were carefully filled with a de-aired solution of water with a precisely known salinity.
The open end of the expansion tube was placed within the scale beaker below the water level, thereby creating a closed system that is able to accommodate any volume changes in the tank.
The mixing pump was switched on to keep the ambient fluid well-mixed.
Both of the constant temperature sidewalls were set to approximately 2$^{\circ}$C above the estimated freezing temperature for the particular ambient salinity, and the system was left to come to thermal equilibrium, which typically involved a slight contraction of volume in the tank and an associated drawing of fluid from the scale beaker.
Over time, the mass measured by the scale adjusts to become steady, at which time the experiment is ready to begin.

The freezing plate temperature is lowered to a desired point below the ambient freezing temperature, initiating ice growth on the plate.
The ambient plate temperature is maintained just above the ambient freezing temperature.
These thermal boundary conditions were maintained for a period of over 24\,hours, during which time the ambient temperature and beaker mass were logged, and high-resolution images from the Basler AG camera recorded at 30\,minute intervals.
After this $\sim$24\,hour logging period, the freezing plate temperature was returned to that of the ambient plate and the ice was allowed to melt back into the mixed ambient fluid.
Once the ambient fluid returns to thermal equilibrium, the next experiment is commenced with a different prescribed freezing plate temperature and/or tank orientation.
The ambient fluid was completely replaced between experiments with different initial ambient salinities.


\subsection{Parameter Space}

The freezing plate temperatures $T_{FP}$ explored here are $-10^{\circ}$C, $-15^{\circ}$C, and $-20^{\circ}$C, which are all warmer than the eutectic temperature of salty water at which compositional mushy layers form ($T_{e}=-21.1^{\circ}$C).
The initial ambient salinities employed here are approximately 17, 33, 48, 73, and 102\,g/kg.
For initial salinities greater than 45\,g/kg, the warmest freezing temperature setting ($T_{FP}=-10^{\circ}$C) does not produce sufficiently detectable ice, so only freezing temperatures of $T_{FP}=-15^{\circ}$C and $T_{FP}=-20^{\circ}$C are used.
For the highest salinity experiments (102\,g/kg), ice produced in the bottom orientation was too buoyant and would detach from the freezing plate and float up through the ambient and melt, so only the top and side orientations are included here.
This approach provides a set of 34 different experiments, for which the initial ambient salinities, freezing temperature settings, and tank orientations are depicted in Figure \ref{fig:parameter_space}.


\subsection{Analysis: Calculating the Bulk Solid Mass Fraction}

The quantitative diagnostics here are the logs of the ambient fluid temperature and beaker mass, and the high-resolution images, which combine to provide estimates of the bulk solid mass fraction $\phi$ of the mushy layer.
The solid mass fraction is defined by the ratio of the mass of solid ice to the total mass of the mushy layer,
\begin{equation}
\phi = \frac{m_{s}}{m_{b}+m_{s}},
\label{eqn:SolidFraction}
\end{equation}
where $m_{s}$ and $m_{b}$ refer to mass of solid ice and liquid brine (kg), respectively.
A solid mass fraction of $\phi=1$ represents solid ice with no liquid brine, and a value of $\phi=0$ is entirely liquid brine without any solid ice.
Note that the solid mass fraction $\phi$ and the brine mass fraction $\phi_{l}$ ($\phi_{l}=m_{b}/\left(m_{b}+m_{s}\right)$), which is often used in sea ice models in place of the solid mass fraction \citep[e.g.,][]{hunke_etal2015}, sum to unity; $\phi + \phi_{l} = 1$.
The definition above describes a bulk solid mass fraction which is useful as it provides a domain average of the varying fine scale structures, such that the large-scale properties of the mushy layer can be characterised.
Here, we follow \cite{wettlaufer_etal1997} by employing two distinct approaches to estimate the bulk solid mass fraction in our experiments; these are based on the conservation of salt and the conservation of mass, respectively, and are described below.
Note that in reality the solid mass fraction exhibits variability within the sea ice, which is not able to be captured with our bulk conservation methods employed here.

\subsubsection{Bulk Salt Conservation Method}

The conservation of salt requires the total amount of salt in the system to remain constant throughout the experiment, that is, 
\begin{equation}
    \underbrace{S_{o}\rho_{o} V_{t} \rule[-2pt]{0pt}{5pt}}_{\mbox{Initial Salt}} =
    \underbrace{(1-\phi)S_{b}\rho_{b} V_{m}  \rule[-2pt]{0pt}{5pt}}_{\mbox{Brine Inclusions}}
    + \underbrace{S_{a}\rho_{a}(V_{t}-V_{m}) \rule[-2pt]{0pt}{5pt}}_{\mbox{Ambient Fluid}}
    + \underbrace{S_{e}m_{e} \rule[-2pt]{0pt}{5pt}}_{\mbox{Expelled Fluid}},
    \label{eqn:SoluteConDerv}
\end{equation}
where $S_{o}$, $S_{b}$, $S_{a}$, and $S_{e}$ refer to the salinities of the initial ambient fluid, brine inclusions, ambient fluid during ice growth, and expelled fluid in the beaker, respectively (g/kg); $\rho_{o}$, $\rho_{b}$, and $\rho_{a}$ are the densities of the initial ambient fluid, brine inclusions, and ambient fluid during ice growth, respectively (kg/m$^3$); $V_{t}$ and $V_{m}$ refer to the volumes of the tank and mushy layer, respectively (m$^3$); and $m_{e}$ is the mass of ambient fluid expelled into the beaker (kg).

Rearranging Equation \ref{eqn:SoluteConDerv} for the solid mass fraction produces,
\begin{equation}
\phi_{S} = \frac{S_{o}\rho_{o}V_{t}-S_{b}\rho_{b}V_{m}-S_{a}\rho_{a}(V_{t}-V_{m})-S_{e}m_{e}}{-S_{b}\rho_{b}V_{m}},
\label{eqM:Solid Frac Solute}
\end{equation}
where we use the subscript $S$ identify this solid mass fraction estimate as that provided by the salt conservation method.

\subsubsection{Bulk Mass Conservation Method}

The conservation of mass provides a mass balance model for the solid mass fraction $\phi$ where the mass of the expelled fluid is directly measured, and the masses of the initial system, solid ice, brine inclusions, evolving ambient fluid are estimated by the products of their respective densities and volumes.
That is,
\begin{equation}
    \underbrace{\rho_{o} V_{t} \rule[-2pt]{0pt}{5pt}}_{\mbox{Initial Mass}} =
    \underbrace{\phi \rho_{ice} V_{m} \rule[-2pt]{0pt}{5pt}}_{\mbox{Solid Ice}} +
    \underbrace{(1-\phi)\rho_{b} V_{m} \rule[-2pt]{0pt}{5pt}}_{\mbox{Brine Inclusions}}
    + \underbrace{\rho_{a}(V_{t}-V_{m})\rule[-2pt]{0pt}{5pt}}_{\mbox{Ambient Fluid}}
    + \underbrace{m_{e} \rule[-2pt]{0pt}{5pt}}_{\mbox{Expelled Fluid}},
    \label{eqn:MassConDerv}
\end{equation}
where $\rho_{o}$, $\rho_{ice}$, $\rho_{a}$ and $\rho_{b}$ are the densities of the initial ambient fluid, solid ice, ambient fluid during ice growth, and brine channel fluid, respectively (kg/m$^3$); $V_{t}$ and $V_{m}$ refer to the volumes of the tank and mushy layer, respectively (m$^3$); and $m_{e}$ is the mass of ambient fluid expelled into the beaker (kg).
Equation \ref{eqn:MassConDerv} can then be rearranged for $\phi$, to provide the mass conservation method for estimating the solid mass fraction of mushy layer, that is,
\begin{equation}
\phi_{M} = \frac{\rho_{o}V_{t}-m_{e}-\rho_{a}(V_{t}-V_{m})-\rho_{b}V_{m}}{(\rho_{ice}-\rho_{b})V_{m}},
\label{eqn:SolidFracMass}
\end{equation}
where we use the subscript $M$ to identify this solid mass fraction estimate as that provided by the mass conservation method.

\subsubsection{Quantifying Terms for the Conservation Methods}

\begin{table}[!ht]
\caption{Terms needed for the Conservation Methods}
\begin{center}
\begin{tabular}{|c|c|c|c|}
\underline{Term} & \underline{Units} & \underline{Description} & \underline{Provenance} \\
$h$ & m & mushy layer thickness & estimated from photos \\
$h_{o}$ & m & initial mushy layer thickness & estimated from photos \\
$h_{f}$ & m & final mushy layer thickness & estimated from photos \\
$V_{t}$ & m$^3$ & volume of tank & directly measured \\
$V_{m}$ & m$^3$ & volume of mushy layer & needs $h$ \\
$m_{e}$ & kg & mass of expelled fluid & directly measured and logged \\
$\rho_{e}$ & kg/m$^3$ & density of expelled fluid & directly measured \\
$S_{e}$ & g/kg & salinity of expelled fluid & Eqn. \ref{eqn:S_from_rho}; needs $\rho_{e}$ \\
$\rho_{o}$ & kg/m$^3$ & initial ambient fluid density & directly measured \\
$S_{o}$ & g/kg & initial ambient fluid salinity & Eqn. \ref{eqn:S_from_rho}; needs $\rho_{o}$ \\
$\rho_{ice}$ & kg/m$^3$ & density of solid ice & Eqn. \ref{eqn:rho_ice}; needs $T_{ice}$ \\
$T_{ice}$ & $^\circ$C & interior temperature of solid ice & Eqn. \ref{eqn:temp_ice}; needs $T_{liq}$ \\
$T_{liq}$ & $^\circ$C & liquidus temperature & Eqn. \ref{eqn:liquidus}; needs $S_{a}$ \\
$S_{b}$ & g/kg & salinity of brine inclusions & Eqn. \ref{eqn:sal_brine}; needs $T_{ice}$ \\
$\rho_{b}$ & kg/m$^3$ & density of brine inclusions & Eqn. \ref{eqn:S_from_rho} \\
$\rho_{f}$ & kg/m$^3$ & final ambient fluid density & directly measured \\
$S_{f}$ & g/kg & final ambient fluid salinity & Eqn. \ref{eqn:S_from_rho}; needs $\rho_{f}$ \\
$S_{a}$ & g/kg & salinity of the ambient fluid & Eqn. \ref{eqn:amb_sal}; needs $S_{o}$, $S_{f}$, $h$, $h_{o}$, $h_{f}$  \\
$T_{a}$ & $^\circ$C & temperature of the ambient fluid & directly measured and logged \\
$\rho_{a}$ & kg/m$^3$ & density of the ambient fluid & Eqn. \ref{eqn:S_from_rho}; needs $S_{a}$ \\
\end{tabular}
\end{center}
\label{default}
\end{table}%

The mass of the fluid expelled into the beaker, $m_{e}$, was logged in time for the duration of each experiment.
The salinity of the expelled fluid $S_{e}$ was calculated from direct measurements of its density with a precision Anton-Paar densimeter and the equation of state for NaCl and fresh tapwater given by \cite{notz2005},
\begin{equation}
\rho = 998.43 + 0.69722S + 2.5201\times10^{-4}S^{2},
\label{eqn:S_from_rho}
\end{equation}
which was intended for salinities up to $S=260$\,g/kg and temperatures between 0$^\circ$C and -20$^\circ$C from the data of \cite{weast1971}; note that for this temperature range, the maximum error from neglecting the temperature dependence of the density is 2\% \citep{notz2005}.

The volume of the tank $V_{t}$ is constant and determined from accurate measurements of the tank geometry.
The initial ambient fluid density $\rho_{o}$ is directly measured with a precision Anton-Paar densimeter.
The initial ambient fluid salinity $S_{o}$ is calculated from its density with Eqn. \ref{eqn:S_from_rho}.

Following \cite{notz2005}, we estimate the density of the solid ice $\rho_{ice}$ with the equation of state given by \cite{pounder1965},
\begin{equation}
\rho_{ice} = 916.8 - 0.1403T_{ice},
\label{eqn:rho_ice}
\end{equation}
where $T_{ice}$ is the solid ice temperature in $^\circ$C.
As we do not have direct measurements of temperature within the solid ice, we approximate it as the mid-point between the freezing plate temperature $T_{FP}$ and the ambient fluid--ice interface temperature $T_{liq}$.
The interface temperature $T_{liq}$ is assumed to be the liquidus temperature for water with the ambient salinity $S$, which is given by \cite{weast1971} as,
\begin{equation}
T_{liq}(S) = -5.92\times10^{-2}S - 9.37\times10^{-6}S^{2} - 5.33\times10^{-7}S^{3},
\label{eqn:liquidus}
\end{equation}
for the range of $S$ up to 230\,g/kg.
Thus,
\begin{equation}
T_{ice}(S) = \frac{T_{FP}+T_{liq}(S)}{2},
\label{eqn:temp_ice}
\end{equation}
which is used in Eqn. \ref{eqn:rho_ice} to give us the density of solid ice $\rho_{ice}$.

The thickness of the mushy layer $h$ increases as the ice grows; this thickness was estimated from the high-resolution images captured throughout the experiments.
The light intensity discontinuity across the mushy layer and liquid ambient boundary was identified with image analysis software and used to calculate the tank-average thickness of the mushy layer.
This mushy layer thickness was multiplied by the relevant geometry of the tank to obtain an estimate of the mushy layer volume $V_{m}$.
This simple approach assumes there is no three-dimensional structure in the mushy layer growth, which in reality is not the case.
To account for three-dimensionality in the mushy layer, a small geometrical volume correction is applied (described below).
Note that the initial mushy layer thickness $h_{o}$ is not necessarily zero; the first photo of the experiment is typically 30 minutes after the freezing plate temperature has been adjusted.
The final mushy layer thickness $h_{f}$ is the thickness of the mushy layer when it has reached an equilibrium, which was defined as the time when the average mushy layer thickness has not increased by more than 0.5\% relative to the previous image.

The salinity of the brine inclusions $S_{b}$ is assumed to be the salinity corresponding to the boundary between mushy layer and aqueous solution for that particular ice temperature, which is given by the liquidus curve.
That is, the equation for the liquidus (Eqn. \ref{eqn:liquidus}) can be inverted to give brine salinity as a function of ice temperature ($^\circ$C),
\begin{equation}
S_{b} = -17.6T_{ice}-0.389T_{ice}^2-3.62\times10^{-3}T_{ice}^3.
\label{eqn:sal_brine}
\end{equation}
With the brine salinity we can then estimate the brine density with Eqn. \ref{eqn:S_from_rho}.
Note that this equation for brine salinity as a function of ice interior temperature is specific for aqueous solutions of NaCl and water, and differs from that employed by sea ice models \citep{hunke_etal2015},
\begin{equation}
S_{b} = \left({10^{-3}-\frac{0.054}{T_{ice}}}\right)^{-1},
\label{eqn:cice_brineS}
\end{equation}
which is specific for brine salinity in sea ice formed from actual seawater.

The final salinity of the ambient fluid $S_{f}$ is calculated from the final density of the ambient fluid $\rho_{f}$, which is measured directly with a precision Anton-Paar densimeter, and Equation \ref{eqn:S_from_rho}.
The evolving ambient fluid density $\rho_{a}$, however, is not measured during the experiment, so the evolving salinity of the ambient fluid $S_{a}$ needs to be estimated using measurements of the initial and final ambient fluid salinities, $S_{o}$ and $S_{f}$, respectively.
The salinity of the ambient fluid evolves due to the growth of the mushy layer and the brine rejection from the mushy layer; these processes are not constant in time, so approximating the ambient fluid salinity with a linear evolution between the initial and final salinities is not valid.
Thus, we follow the approach of \cite{wettlaufer_etal1997}, and relate the evolution of the ambient salinity to the evolution of the mushy layer thickness $h$ by way of a scaling factor; that is,
\begin{equation}
H_{c} = \frac{h(t) - h_{o}}{h_{f} - h_{o}},
\label{eqn:height_scaling}
\end{equation}
where $h(t)$ is the mushy layer thickness at time $t$, and $h_{o}$ and $h_{f}$ are the initial and final mushy layer thicknesses.
The estimated salinity of the ambient fluid is then given by,
\begin{equation}
S_{a}(t) = H_{c}\left(S_{f} - S_{o}\right) + S_{o}.
\label{eqn:amb_sal}
\end{equation}
The temperature of the ambient fluid $T_{a}$ is directly measured and logged throughout the experiment.
The evolving density of the ambient fluid $\rho_{a}$ is then able to be calculated from the ambient fluid temperature and salinity with an appropriate equation of state, for which we use Equation \ref{eqn:S_from_rho}.

In summary, the mass and salt conversation methods require the knowledge of several salinities, temperatures, densities, masses, volumes, and thicknesses.
Some are directly measured ($V_{t}$, $m_{e}$, $\rho_{e}$, $\rho_{o}$, $\rho_{f}$, $T_{a}$), some are estimated from high-resolution photos ($h$, $h_{o}$, $h_{f}$, $V_{m}$), and some are approximated from established and/or adapted relationships ($S_{e}$, $S_{o}$, $\rho_{ice}$, $T_{ice}$, $T_{liq}$, $S_{b}$, $\rho_{b}$, $S_{f}$, $S_{a}$, $\rho_{a}$).

\section{Results}

\begin{figure}[!ht]
\begin{center}
\includegraphics[width=1.0\textwidth]{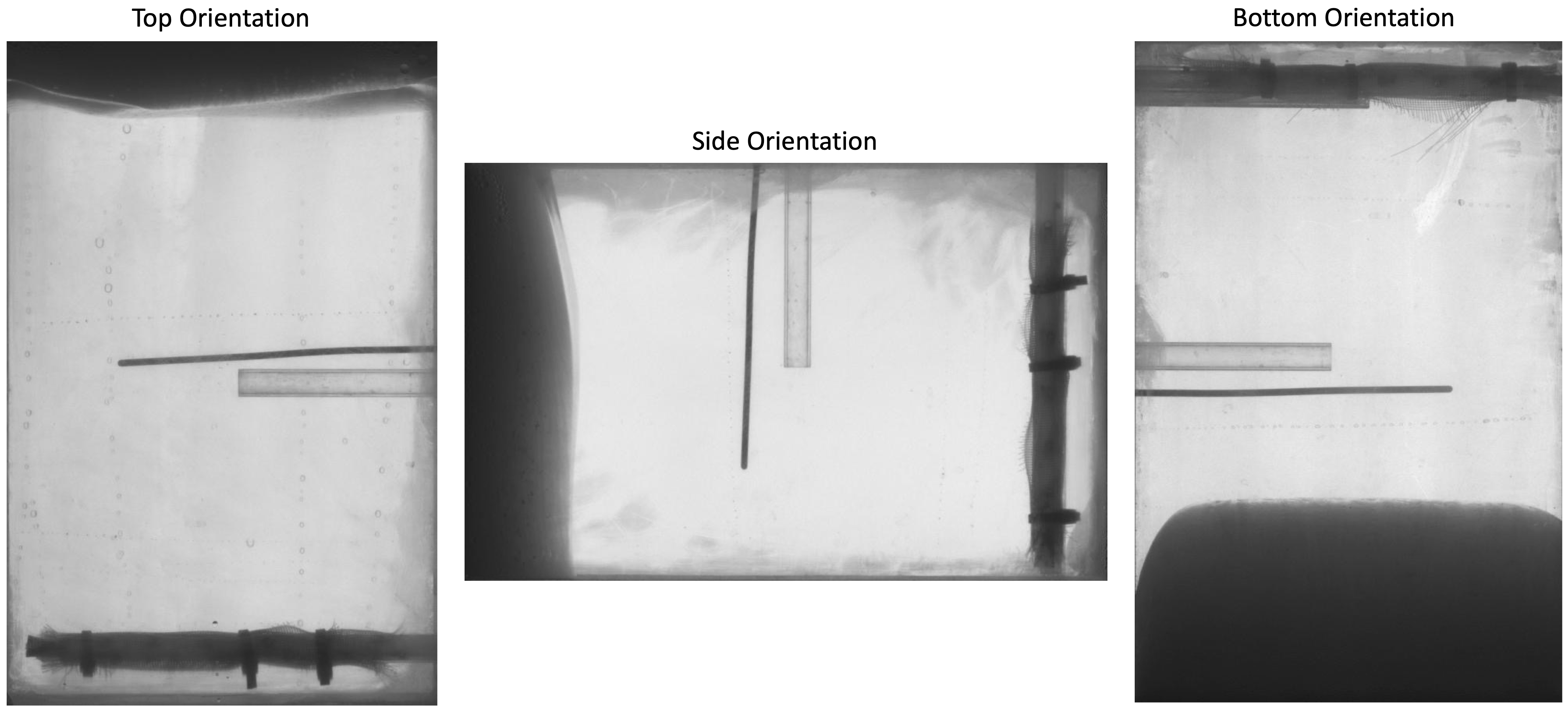}
\caption{Photos of the equilibrated mushy layer for the top, side, and bottom freezing plate orientations (left to right) for experiments with initial ambient salinities of $S_{o}=33$\,g/kg and freezing plate temperatures of $T_{FP}=-20^{\circ}$C.}
\label{fig:exp_pics}
\end{center}
\end{figure}

Once the experiments began, the mushy layer was observed to form on the freezing plate within the first 30\,mins.
The initial distribution of the mushy layer was uniform across the freezing plate, indicative of a uniform boundary condition imposed by the heat exchanger.
In time, the edges of the mushy layer within $\sim$5\,mm of the tank sidewalls and ice face became rounded, suggesting a small amount of unintentional heat entering the tank from the laboratory; for the bottom oriented case, this rounded mushy layer edge was more obvious (Fig. \ref{fig:exp_pics}).
Also, when estimating the mushy layer volume from the high-resolution photos, a minor geometric correction is applied to account for the mushy layer having rounded edges.
As the mushy layer continues to grow, the ice face tended to develop low-mode three-dimensional structures in the plane parallel to the freezing plate (vertical and into the page in Fig. \ref{fig:exp_schematic}).
The exact nature of these structures appears to depend on minor differences in initial conditions since repeating experiments didn't necessarily reproduce identical shapes of mushy layer face structures.
Experiments with bottom oriented freezing plates exhibited relatively less three-dimensionality to the mushy layer face, suggesting the structure is perhaps related to the gravity drainage process in the top and side orientations.
Nevertheless, the amplitude of these features were small relative to the mushy layer thickness; as such, the mushy layer thickness and volume estimates were developed by taking a tank average of all ice edges visible.
The detection limit of this method is given by the thickness per pixel, which is approximately 0.7\,mm/pixel.
Considering the complications associated with the rounded edges and three-dimensionality corrections, generous uncertainties (up to 5\%, depending on the extent of the rounded edges and three-dimensionality) are applied to the mushy layer volume estimates.

\begin{figure}[!ht]
\begin{center}
\includegraphics[width=1.0\textwidth]{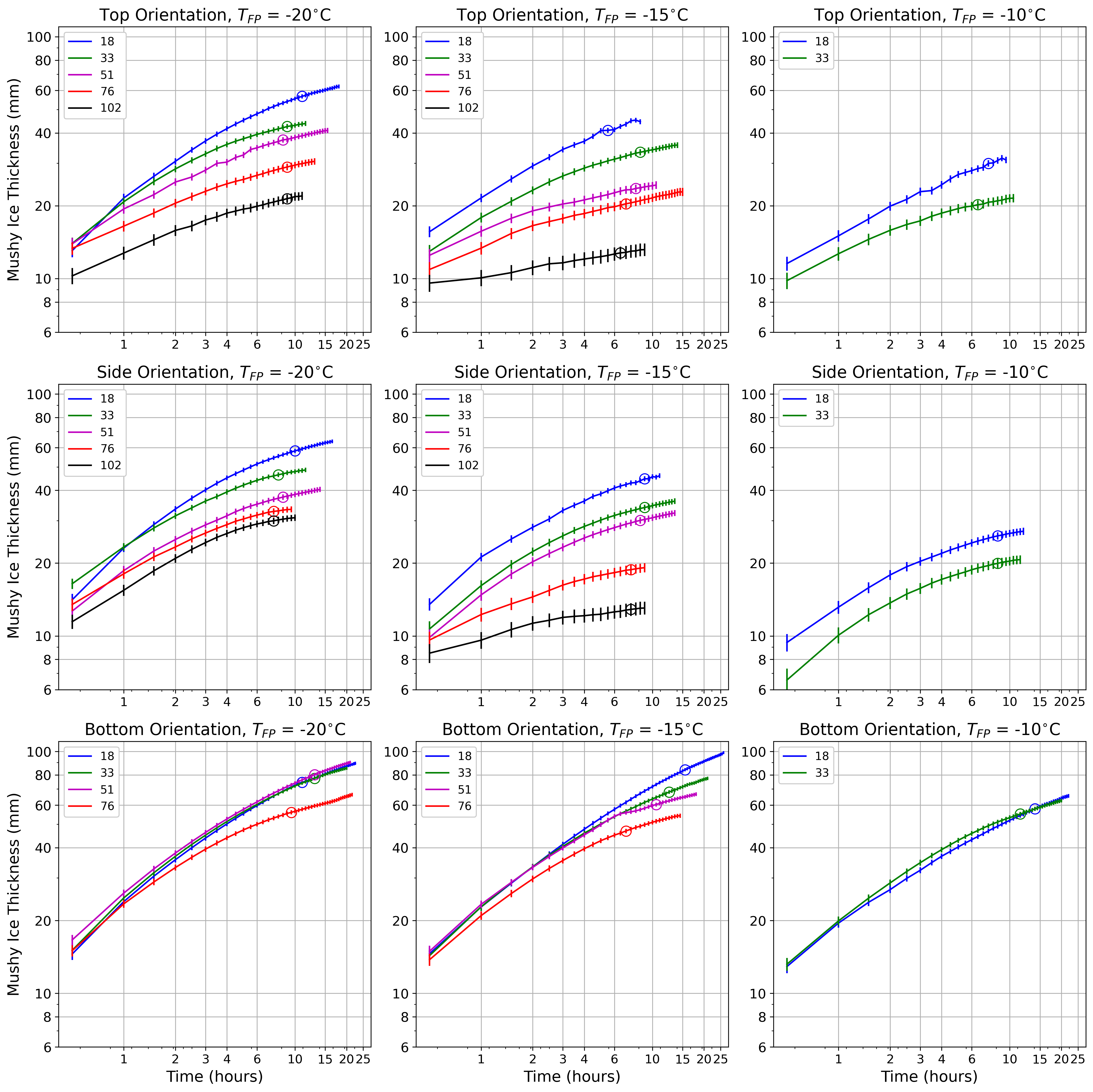}
\caption{The time evolution of bulk mushy layer thickness for all experiments; the columns indicate the different freezing plate temperatures increasing from left to right, and the rows indicate the different freezing plate orientations. The line colours represent the different initial ambient salinities (see legend; in g/kg). The vertical dashes indicate the times that photos were taken, and their extent is indicative of the measurement uncertainty. The circles represent the bulk mushy layer thickness when the experiment has reached equilibrium.}
\label{fig:bulk_thickness}
\end{center}
\end{figure}

Analysis of the high resolution photos allows us to quantify the bulk mushy layer thickness and how it changes over the course of an experiment (Fig. \ref{fig:bulk_thickness}).
For all experiments, the mushy layer thickness monotonically increases in time, with the rate of thickness change tending to decrease in time.
The responses of the mushy layer thickness to the initial ambient salinity and freezing plate temperatures are intuitive and consistent for the top and side orientations; experiments with higher initial salinities and warmer freezing plate temperatures exhibit reduced mushy layer thicknesses and growth rates.
The bottom orientation experiments exhibit less sensitivity to the initial ambient salinity and freezing plate temperature, and faster growth rates.

\begin{figure}[!ht]
\begin{center}
\includegraphics[width=1.0\textwidth]{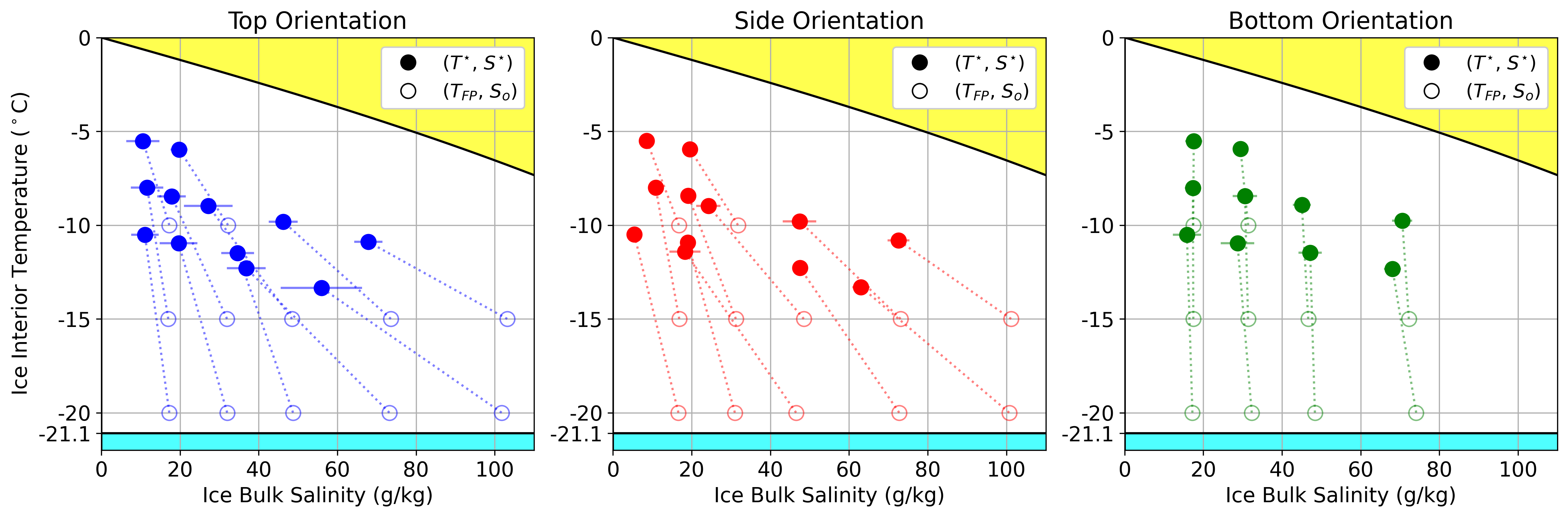}
\caption{The initial (hollow circles) and final (solid circles) ice bulk salinity and interior temperature for the top, side, and bottom freezing plate orientation (left to right). The horizontal lines through the solid circles represent the range of final ice bulk salinities based on the two methods to calculate the solid mass fraction. The yellow, white, and blue regions represent the aqueous solution, mushy layer, and compositional mushy layer regimes, respectively, as per Figure \ref{fig:parameter_space}.}
\label{fig:ice_conditions}
\end{center}
\end{figure}

Figure \ref{fig:ice_conditions} shows the evolution of the interior ice conditions.
The ice interior temperature characterises the bulk thermal conditions of the ice.
The ice bulk salinity reflects the ratio of the total mass of salt contained in brine within the mushy layer (in grams) to the total mass of the mushy layer and brine mixture (in kilograms); this is the salinity of the aqueous solution that results from the melting of the mushy layer.
Here, the initial ice bulk salinity is assumed to be that of the initial ambient fluid salinity $S_{o}$, and the initial ice interior temperature is that of the freezing plate $T_{FP}$.
The final ice interior temperature $T^{\star}$ is given by Eqn. \ref{eqn:temp_ice}, where we use the freezing plate temperature $T_{FP}$, and the liquidus temperature $T_{liq}$ based on the final ambient salinity $S_{f}$.
The final ice bulk salinity $S^{\star}$ is calculated using the solid mass fraction $\phi$ and the brine salinity $S_{b}$, which is given by Eqn. \ref{eqn:sal_brine} using the final ice interior temperature $T^{\star}$; the relationship between the ice bulk salinity, brine salinity, and solid mass fraction is,
\begin{equation}
\phi = 1-\frac{S^{\star}}{S_{b}}\quad \implies \quad S^{\star} = \left(1 - \phi\right)S_{b}.
\label{eqn:phi_S_Sb}
\end{equation}
Here for $\phi$ we use the average value of $\phi_{M}$ and $\phi_{S}$, and provide an indication of the differences of ice bulk salinities estimated by the two different methods (horizontal lines through the final ice bulk salinities).

The final ice interior temperature is, by definition, always warmer than the initial ice interior temperature.
The final ice bulk salinity is fresher that the initial ice bulk salinity for the top and side orientations; for the bottom orientation, however, the initial and final ice bulk salinities are virtually the same.
This freshening of the ice bulk salinity reflects the gravity drainage of salty brine from the mushy layer in the top and side orientations, which is a process that is not able to occur in the bottom orientation experiments.

\begin{figure}[!ht]
\begin{center}
\includegraphics[width=1.0\textwidth]{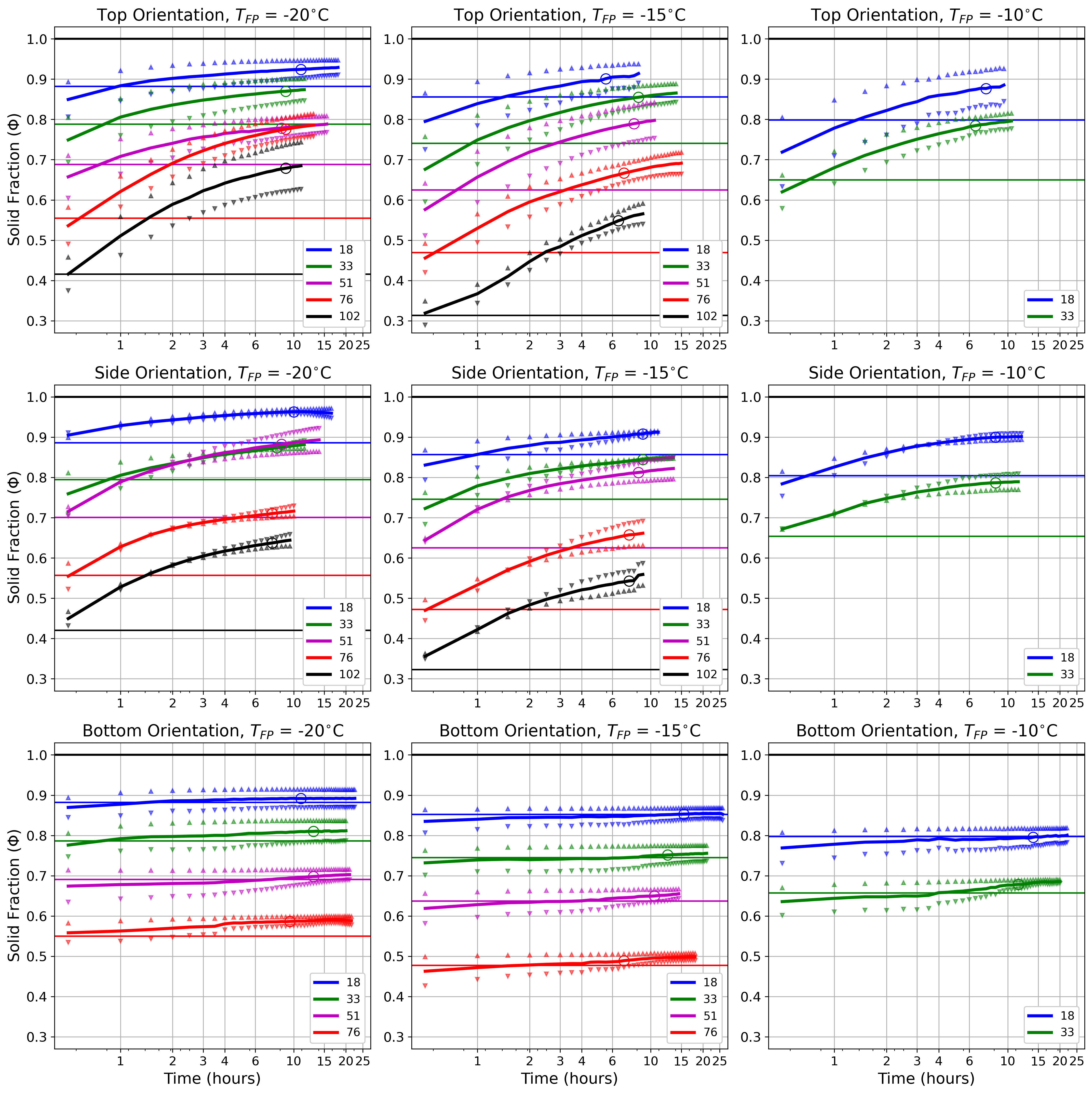}
\caption{The time evolution of the solid mass fraction for all experiments; the columns indicate the different freezing plate temperatures increasing from left to right, and the rows indicate the different freezing plate orientations. The line colours represent the different initial ambient salinities (see legend; in g/kg). The upright triangles are the solid mass fractions calculated with the salt conservation method $\phi_{S}$; the inverted triangles are the solid mass fractions calculated with the mass conservation method $\phi_{M}$; the lines follow the average value of the two. The circles represent the solid mass fraction when the experiment has reached equilibrium. The horizontal lines indicate the predicted solid mass fraction $\phi$ based on the initial ice bulk salinity (approximated by $S_{o}$) and interior ice temperature $T_{ice}$ (Eqns. \ref{eqn:sal_brine} \& \ref{eqn:phi_S_Sb}).}
\label{fig:phi_time}
\end{center}
\end{figure}

The solid mass fractions calculated by the mass and salt conservation methods over the course of each experiment are shown in Figure \ref{fig:phi_time}.
In general, the solid mass fractions decrease for increasing ambient salinities, and warming freezing plate temperatures.
The experiments with top and side orientations exhibit solid mass fractions that increase in time, and the rate at which these solid mass fractions increase tends to increase with ambient salinity.
The bottom orientation experiments tend to remain near their initial solid mass fractions.

The initial ice bulk salinity (approximated by the initial ambient fluid salinity $S_{o}$) and the ice interior temperature $T_{ice}$ can be used with Equations \ref{eqn:sal_brine} \& \ref{eqn:phi_S_Sb} to calculate a predicted solid mass fraction, which are shown in Figure \ref{fig:phi_time} as horizontal lines.
These predicted solid mass fractions exhibit good agreement with the measured $\phi$ during early stages, and throughout for the bottom orientation experiments.
Interestingly, while the top and bottom orientation experiments tend to have larger $\phi_{S}$ relative to $\phi_{M}$, the side orientation experiments have larger $\phi_{M}$; the reason for this is unknown.

\begin{figure}[!ht]
\begin{center}
\includegraphics[width=1.0\textwidth]{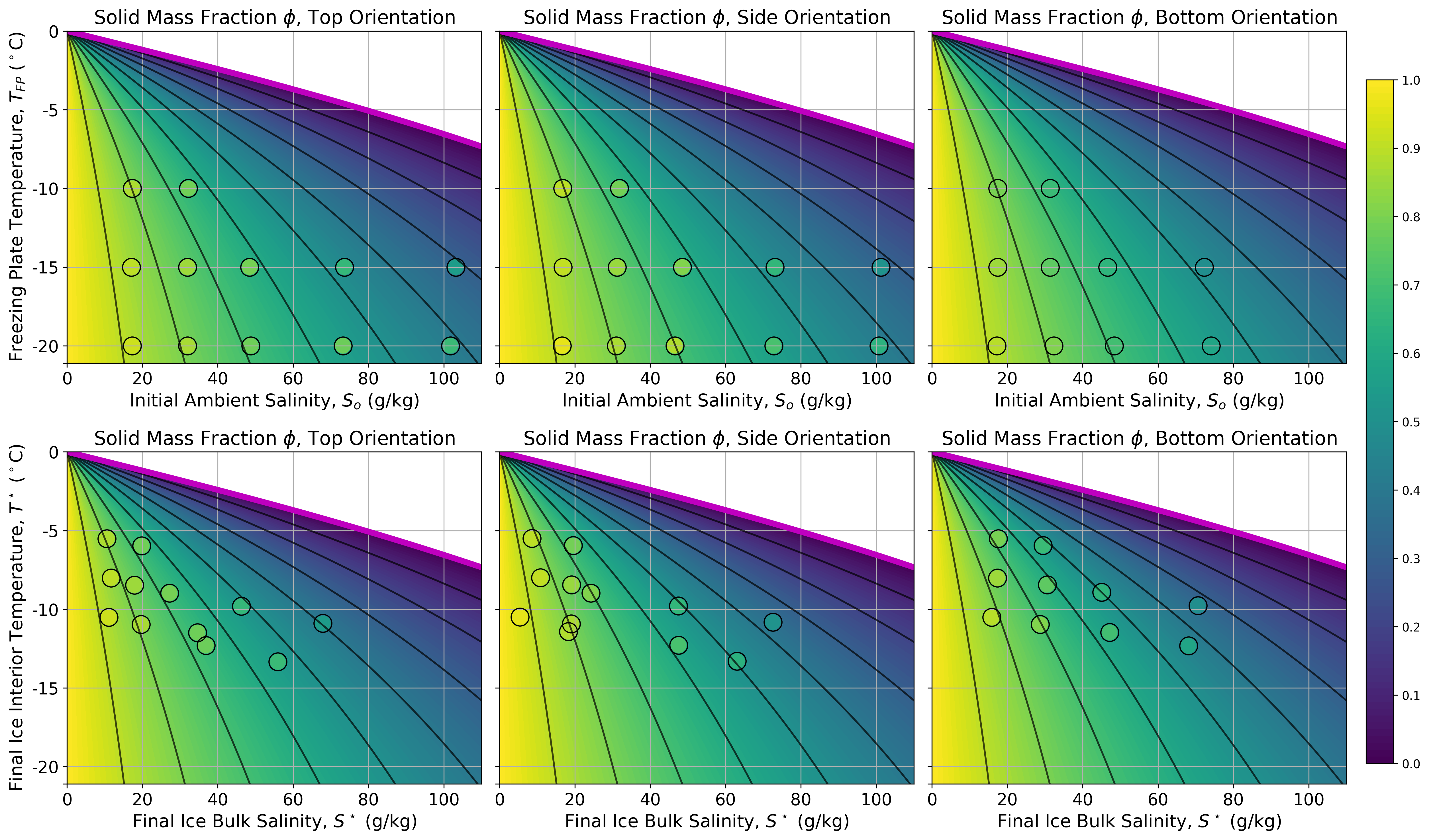}
\caption{Measured solid mass fractions at equilibrium plotted by their respective freezing plate temperatures $T_{FP}$ and initial ambient salinities $S_{o}$ (top row), and their respective final ice interior temperatures $T^{\star}$ and bulk salinities $S^{\star}$ (bottom row). The background colourmap indicates the solid mass fractions predicted by Equations \ref{eqn:liquidus}, \ref{eqn:temp_ice}, \ref{eqn:sal_brine} \& \ref{eqn:phi_S_Sb} using a range ice bulk salinities and interior temperatures. The contours are at $\phi=0.1$ intervals, and the magenta line represents the liquidus curve.}
\label{fig:phi_w_BS_T}
\end{center}
\end{figure}

The equilibrated solid mass fractions are sensitive to the salinity and temperature conditions of the system.
Figure \ref{fig:phi_w_BS_T} shows the measured solid mass fractions of the experiments plotted by their respective initial ambient salinities $S_{o}$ and freezing plate temperatures $T_{FP}$ (top row), and their final ice bulk salinities $S^{\star}$ and interior temperatures $T^{\star}$ (bottom row).
These plots include the solid mass fractions predicted by Equations \ref{eqn:liquidus}, \ref{eqn:temp_ice}, \ref{eqn:sal_brine} \& \ref{eqn:phi_S_Sb} for a range of ambient salinities and freezing plate temperatures (top row) and ice bulk salinities and /interior temperatures (bottom row).

In general, there is good agreement between the measured and predicted solid mass fractions; they exhibit consistent behaviours in terms of their relative sensitivities to the salinity and temperature conditions.
The agreement is improved when using the final ice bulk salinities and interior temperatures instead of the initial ambient salinities and freezing plate temperatures.
The agreement between the measured and predicted solid mass fractions is best for experiments with smaller ice bulk salinities; this makes good physical sense as the solid mass fraction converges to $\phi=1$ as system freshens.
The experiments with bottom freezing plate orientations exhibit the best agreement for the initial conditions; this reflects the fact that brine drainage does not occur in this configuration, such that the equilibrium ice conditions are well represented by the known initial conditions.
We hypothesise that the leading sources of the minor disagreement between the measured and predicted is uncertainty in regards to the final ice interior temperature (which we assume to be the mid-point between the freezing plate and liquidus temperatures; Eqn. \ref{eqn:temp_ice}), and variations throughout the mushy layer that are not well represented by the bulk approach employed here.

\begin{figure}[!ht]
\begin{center}
\includegraphics[width=1.0\textwidth]{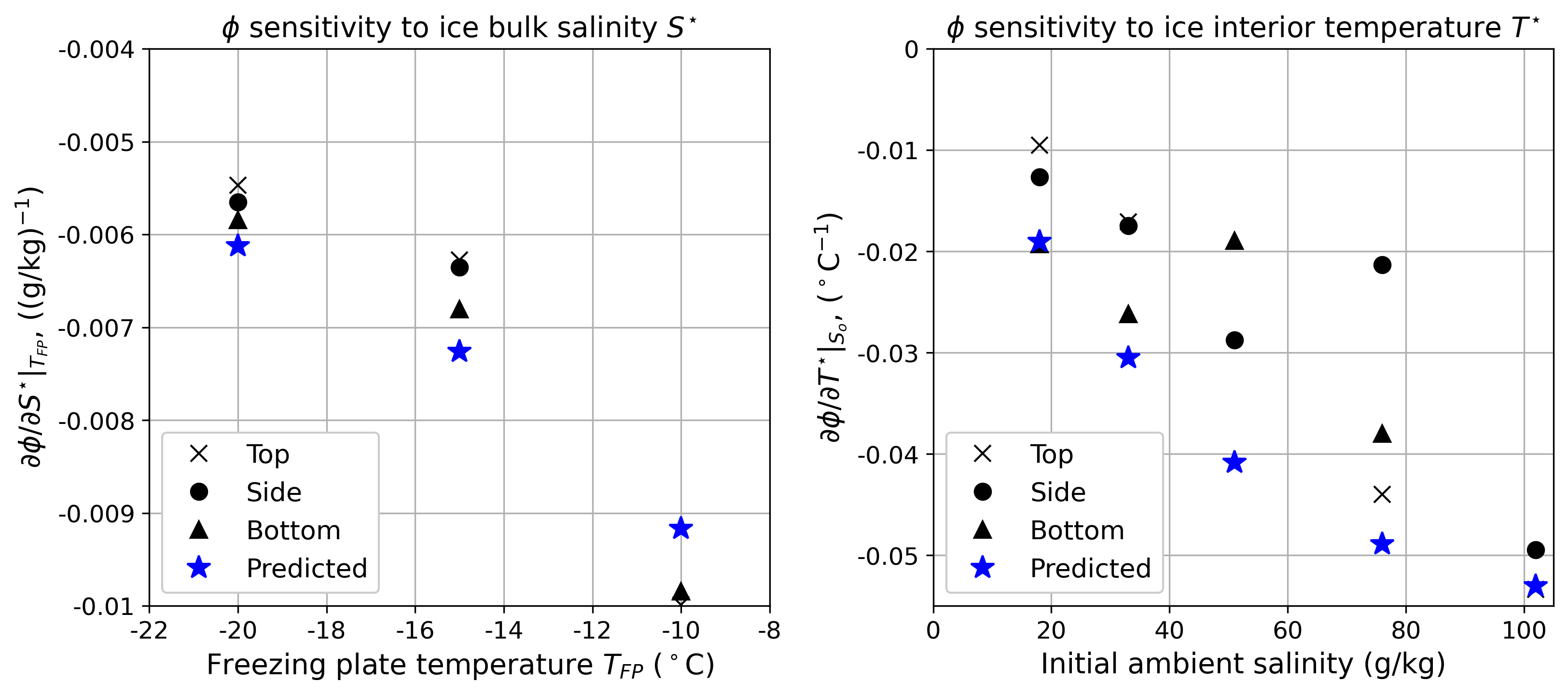}
\caption{The sensitivities of measured solid mass fraction to the ice bulk salinity $\partial\phi/\partial{S^{\star}}|_{T_{FP}}$ (left) and the ice interior temperature $\partial\phi/\partial{T^{\star}}|_{S_{o}}$ (right) for given freezing plate temperatures and initial ambient salinities, respectively. The predicted sensitivities given by the relationships in Equations \ref{eqn:liquidus}, \ref{eqn:temp_ice}, \ref{eqn:sal_brine} \& \ref{eqn:phi_S_Sb} are also included (blue stars).}
\label{fig:phi_sensitivities}
\end{center}
\end{figure}

The sensitivities of the measured solid mass fractions with respect to the final ice bulk salinities and ice interior temperatures are shown in Figure \ref{fig:phi_sensitivities}, plotted by the freezing plate temperature and initial ambient salinity, respectively.
These are calculated by taking the mean gradient of the measured $\phi$ with respect to bulk ice salinity at the three different freezing plate temperatures ($\partial\phi/\partial{S^{\star}}|_{T_{FP}}$; left panel), and the mean gradient of the measured $\phi$ with respect to ice interior temperature at the five different initial ambient salinities ($\partial\phi/\partial{T^{\star}}|_{S_{o}}$; right panel).
Also included are the predicted sensitivities derived from the solid mass fraction predictions shown in Figure \ref{fig:phi_w_BS_T}; for these we use values of $\partial\phi/\partial{S^{\star}}|_{T_{FP}}$ at $S_{o}=33$\,g/kg for the three freezing plate temperatures, and the mean values of $\partial\phi/\partial{T^{\star}}|_{S_{o}}$ at $T_{FP}=\left(-20,-15,-10\right)^{\circ}$C for the five different initial ambient salinities.

The solid mass fraction sensitivity to ice bulk salinity has a strong dependence on the freezing plate temperature; for a given freezing plate temperature, the rate of change of solid mass fraction with respect to ice bulk salinity nearly doubles from $\partial\phi/\partial{S^{\star}}|_{T_{FP}}\approx-0.005$\,(g/kg)$^{-1}$ at $T_{FP}=-20^{\circ}$C to more than $\partial\phi/\partial{S^{\star}}|_{T_{FP}}\approx-0.009$\,(g/kg)$^{-1}$ at $T_{FP}=-10^{\circ}$C.
This is reflected in the convergence of the predicted solid mass fraction contours for warming freezing plate temperatures in Figure \ref{fig:phi_w_BS_T}; this convergence occurs because $\phi$ goes from $\phi=1$ at zero salinities to $\phi=0$ at the liquidus salinity, which decreases for increasing temperatures.
The sensitivity of the solid mass fraction to the ice interior temperature depends on the initial ambient salinity; the rate of change of solid mass fraction with ice interior temperature increases by a factor of 5 between initial ambient salinities of $S_{o}\approx18$\,g/kg to $S_{o}\approx102$\,g/kg.
There is good agreement between the measured sensitivities and those predicted by the relationship in Equations \ref{eqn:liquidus}, \ref{eqn:temp_ice}, \ref{eqn:sal_brine} \& \ref{eqn:phi_S_Sb}.

The sensitivity analysis allows us to determine whether the solid mass fraction is more responsive to the temperature or salinity properties of the system.
For the oceanographically realistic experiments with $S_{o}\approx33$\,g/kg, the sensitivity of solid mass fraction to ice interior temperature is approximately $\partial\phi/\partial{T^{\star}}|_{S_{o}}\approx-0.02^{\circ}$C$^{-1}$, or -2.5\% per 1$^{\circ}$C increase.
To achieve an equivalent magnitude change in the solid mass fraction from a change in the ice bulk salinity requires a salinity increase of approximately 4\,g/kg at $T_{FP}=-20^{\circ}$C, or approximately 2\,g/kg at $T_{FP}=-10^{\circ}$C.
When considering the relative ranges of temperature and salinity variabilities at high latitudes, this sensitivity analysis suggests that the solid mass fraction is temperature dominated.
It follows that increases in ice interior temperature will lead to a reduction in the ice solid mass fraction.

\clearpage

\section{Discussion \& Conclusions}

The experiments demonstrate how the mushy layer growth rate and solid mass fraction are affected by the freezing plate temperature, ambient salinity, and freezing orientation.
Our approach has allowed for a thorough exploration of a wide range of thermodynamic conditions and freezing orientations with a single apparatus and common methodology.
The results exhibit good agreement with existing theories for how the solid mass fraction depends on ice bulk salinity and interior temperature \citep[e.g.,][]{wettlaufer_etal1997, notz2005, feltham_etal2006}.
Mushy layer growth rate is largest for cooler freezing plate temperatures, fresher ambient salinities, and bottom oriented freezing configurations.
The equilibrated mushy layer solid mass fractions are largest for cooler freezing plate temperatures and fresher ambient salinities, and smallest for the bottom oriented freezing configuration where brine drainage is unable to occur.
The sensitivity of the solid mass fraction to the ice temperature and salinity is explored, and it is found that for oceanographically-realistic conditions the solid mass fraction is most sensitive to changes in temperature; every 1$^\circ$C increase in freezing temperatures will result in a 2.5\% decrease of solid mass fraction.

The bottom oriented freezing experiments exhibit substantially different mushy layer behaviour and solid mass fraction evolution, which results from the absence of brine drainage.
The top and side oriented freezing experiments exhibit very little difference; they have similar solid mass fraction evolutions, final values, and sensitivities for the oceanographically relevant range of salinities (i.e., $S_{o}=\left(18,33\right)$\,g/kg).
This finding suggests that the introduction of floe size distribution capabilities in modern numerical sea ice models (e.g., CICE6 v6.1.2), and the associated increase in lateral sea ice dynamics, should not require distinct parameterisations for solid mass fractions arising from vertical versus lateral sea ice growth.

The implications of the relatively high thermal sensitivity of solid mass fraction are worthy of closer attention.
Increases in the freezing temperatures of sea ice, which are observed and predicted to continue under realistic climate forcing scenarios \citep[e.g.,][]{post_etal2019}, will reduce the solid mass fraction of sea ice.
This solid mass fraction reduction will decrease the sea ice brine rejection and associated salt flux into the ocean, and reduce the capacity of the sea ice field to freshen the high latitude oceans during melt.
Decreasing the brine rejection salt flux will ultimately lead to a freshening of the high-latitude waters that feed the deep, bottom and abyssal waters of the global oceans, thereby having the potential to reduce the strength of the global thermohaline overturning circulation.
A reduced solid mass fraction equates to an increased sea ice bulk salinity; when this saltier sea ice melts, the surrounding surface waters are freshened less than they would otherwise be with higher solid mass fraction sea ice.
This increase in the high-latitude surface salinity would tend to weaken the surface stratification of these salt stratified waters \citep[i.e., beta-oceans;][]{stewart_haine2016}, and lead to a reduced sea ice growth rate.
All of these implications share a common consequence of tending to reduce the role of sea ice in the climate system.
Investigating the potential feedbacks caused by solid mass fraction variations in coupled ocean--sea ice and climate models is a primary topic for future research.

\section{Acknowledgments}

We wish to thank Angus Rummery and Tony Beasley for the construction of the apparatus and laboratory assistance.

\bibliographystyle{igs.bst}
\bibliography{stewart_etal_jog.bib}

\end{document}